# First Principal Investigations to Explore the Half-metallicity, Structural, Mechanical, and Optoelectronic Properties of Sodium-Based Fluoroperovskites NaYF$_3$ (Y = Sc and Ti) for Applications in Spintronics and Optoelectronics

Saeed Ullah[1,2,*], Uzma Gul[1], Saad Tariq[3], Riaz Ullah[4], Nasir Rahman[5], Essam A. Ali[6], Mudasser Husain[5], Munawar Abbas[1], and Hafeez Ullah[1,2]

[1]*Institute of Physics, Gomal University, Dera Ismail Khan 29220, KP, Pakistan*
[2]*Department of Physics, University of Buner, Swari, Buner 19290, KP, Pakistan*
[3]*Department of Physics, Faculty of Sciences and Technology, University of Central Punjab, Lahore, Pakistan*
[4]*Department of Pharmacognosy, College of Pharmacy, King Saud University Riyadh Saudi Arabia*
[5]*Department of Physics, University of Lakki Marwat, Lakki Marwat 28420, KP, Pakistan*
[6]*Department of Pharmaceutical Chemistry, College of Pharmacy, King Saud University Riyadh Saudi Arabia*
[*]Corresponding author: Saeed Ullah    Email: saeedullah.phy@gmial.com

## Abstract

A theoretical investigation was conducted on Na-based fluoro-perovskites NaYF$_3$ (Y = Sc, Ti) to examine their structural, optical, electronic, and mechanical characteristics for the first time. These cubic compounds exhibit structural stability, maintaining perovskite structures with lattice spacing ranging from 4.15 to 4.26 Å. Computation of elastic parameters confirm their stability, ionic bonding, ductility, and anisotropy. Computed band profiles reveal the half-metallic nature with indirect (M-Γ) bandgaps for the spin-down configurations. Furthermore, density-of-states analysis highlights the role of Y (Sc, Ti) atoms in the metallic character and conduction band contribution. The lack of absorbance in the visible region highlights the materials' suitability for optoelectronic devices. This investigation aims to provide comprehensive insights and encourage further experimental research.

**Keywords:** Half Metallicity; Electronic Properties; Elastic Properties; Fluoro-perovskites; Spintronics, Density of States

## 1. Introduction:

To fulfill the substantial demands of the developing society, modern technical advancements are necessary. The use of novel, cutting-edge materials with enhanced characteristics are therefore strongly favored [1]. Among those materials, the half-metallic compounds are the front-runners and have been widely studied for spintronics applications both experimentally and theoretically [2,3]. Half metallic materials, which were identified in 1980, are the class of materials that display metallic nature for one of the two spin directions while being semiconducting/insulating for the opposite spin [4]. Transition metal oxides, heusler alloys, the diluted magnetic semiconductors (DMS), perovskites, and the double-perovskites are only a few of the compounds that exhibit the phenomena of half-metallicity [5–9]. The spin-dependent features of half-metallic materials offer the possibility for the development of innovative spin-polarized microelectronic devices like the non-volatile magnetic random-access memories (MRAMs) and magnetic sensors [10–12].

Owing to their exceptional qualities, oxide perovskites have extensively investigated for a broad range of industrial applications and have widely been utilized in electrical and microelectronic devices [13,14]. For example, in Ref. [15], the authors reported the stability of magnesium-based manganese oxide ($MgMnO_3$) perovskite synthesized at low temperatures by employing the sol-gel process and demonstrated that a magneto-dielectric effect is evidently present for the temperature below the blocking temperature. Similarly, the lead-based oxide perovskites such as the $Pb(Zr,Ti)O_3$ and $Pb(Mg_{0.33}Nb_{0.67})O_3$-$PbTiO_3$ have been the focus of intense research and have commercially been employed in the electronic equipments as ultrasonic motors, medical transducers, sensors, and dielectric capacitors [16,17]. The environmental and health-related risks associated with these materials are, however, one of their main downsides. These materials roughly contain 70% by weight of hazardous lead oxide, posing a potential threat to the environment [18]. Researchers in the field of materials science are looking forward to focus their efforts on discovering environmentally friendly materials with comparable features [14]. To find an appropriate half-metallic compound for applications in spintronics, materials scientists are recently exploring the lead-free halide-perovskites with the general formula $ABF_3$, where F belongs to the halogen group [5]. Furthermore, these materials provide opportunities for applications, for example, in thermoelectric devices, piezoelectricity, colossal magnetoresistivity, photoluminescence, and high-temperature superconductors [1,19].

Driven by the flourishing promise of halide-perovskites in modern technology, our pursuit is to unveil the physical prospects of NaYF$_3$ (Y = Sc and Ti) through a rigorous investigation of its structural, elastic, electrical, and optical properties.

## 2. Computational Methodology

The full-potential linear augmented plane wave (FP-LAPW) approach [20,21], programmed in the WIEN2K package [22], was employed to work out the Kohn Sham (KS) equations. Being a vital quantum mechanical tool for solving the many-body problems it is one of the trustworthy methods for examining the physical characteristics of the materials [23,24]. The structural and elastic characteristics of studied compounds were ascertained by employing the exchange correlation (X$_c$) potential in the generalized gradient approximation (GGA). The electronic and optical characteristics of the studied compounds were assessed by deploying the Tran and Blaha modified Becke-Johnson potential, which, in contrast to the GGA, does not undervalue the bandgaps of the materials. To extract the equilibrium structural parameters, the energy vs volume optimized curves were fitted to the well-known Birch-Murnaghan equation of state. For convergence to a satisfactory level, with 2000 k-points, the magnitude of the base functions consisting of the plane waves was regulated by the parameters using a value of $RMT \times k_{max}$ equal to 8. In plane wave expansions, $k_{max}$ designates the size of largest $k$-vector while RMT value signifies the minimal radius of the muffin-tin spheres. To eliminate the overlap, each atom was assigned to its own unique RMT value. To ensure consistency, the $G_{MAX}$ value was set to 12 (a.u.)$^{-1}$ while convergence was achieved when the energy threshold value approached 0.001 Ry. The IRelast package built by Jamal Murtaza et al. [25,26] was employed to forecast the elastic constant along with additional mechanical parameters of the studied ternary compounds.

## 3. Results and Discussions

### 3.1 Structural Properties

In this sub-section, we inspect the structural characteristics of the lead-free halide-perovskites NaYF$_3$ (Y = Sc and Ti). Fig. 1 represents the crystal structure of one of the Na-based fluoro-perovskites which crystallizes in simple cubic structure (with space group: $pm\overline{3}m$). As per Wyckoff coordinates, the Na, and Y (Sc and Ti) are positioned at 1a (0, 0, 0) and 1b (0.5, 0.5,

0.5) while fluorine occupies the face center of the unit cell with positions 1c (0, 0.5, 0.5), (0, 0.5, 0.5), or (0, 0.5, 0.5). The Goldschmidt tolerance-factor ($\tau$) was used to assess the structural stability of these ternary compounds [27]:

$$\tau = \frac{r_{Na}+r_F}{\sqrt{2}(r_Y+r_F)} \quad (1)$$

Where $r_{Na}$, $r_Y$, and $r_F$ represents the ionic radii of Na, (Y = Sc and Ti), and F respectively. According to existing literature, materials with a tolerance factor ($\tau$) falling in the range of 0.90 to 1 maintain a stable cubic perovskite structure [28]. Deviations from this range introduce an additional force that significantly alters the material's properties. In the case of the compounds under investigation, the calculated tolerance factors of 0.91 for $NaScF_3$ and 0.927 for $NaTiF_3$ fall within the acceptable range, affirming the structural stability of these studied materials.

We performed structural optimization on the chosen compounds by systematically varying their volumes around the equilibrium cell volume ($V_0$). This volume optimization allowed us to predict various structural characteristics of the compounds under investigation. The energy-versus-volume optimization curves, illustrated in Fig. 2, indicate that as the volume increases, the energy decreases until it reaches the ground state energy value. Beyond this point, as the unit cell volume continues to increase, the energy rises, indicating an approaching instability in the system. To derive essential parameters such as lattice constants, ground state energy and volume, bulk modulus, and the pressure derivative of bulk modulus, we fit the observed curves to the well-established Birch-Muranghan equation of state [29–31]:

$$E(V) = E_0 + \frac{B}{\acute{B}(\acute{B}-1)}\left[V\left(\frac{V_0}{V}\right)^{\acute{B}} - V_0\right] + \frac{B}{\acute{B}(V-V_0)} \quad (2)$$

Table 1 enumerates the extracted equilibrium lattice parameters, and their fitting to the Birch-Muranghan equation has yielded noteworthy results. Notably, $NaTiF_3$ exhibits a significantly higher bulk modulus in comparison to $NaScF_3$. This divergence can be attributed to the disparate electronegativities of titanium (Ti) with a value of 1.54 and scandium (Sc) with a value of 1.35. This distinction implies that a considerably higher pressure is needed to compress the unit cell of $NaTiF_3$, elucidating the observed difference in bulk modulus. The overall tendency of this approximation is supported by the fact that the bulk modulus increases as the lattice constant drops, implying that the estimated outputs are more precise and realistic. Additionally, it is imperative to highlight that the optimized lattice parameters for NaScF$_3$ and NaTiF$_3$, unveiled for the first time within this study, serve as pivotal benchmarks for future investigations concerning these compounds.

**3.2 Elastic Properties:**

The elastic constants (ECs) are essential in describing the mechanical characteristics of the compounds as they explore how these materials respond to external forces. Information regarding toughness and stability of materials can be unveiled by measuring these constants. By computing the components of stress tensor at minute strains and providing the energy in proportion to the lattice strain that preserved volume, the ECs of targeted compounds were determined at zero pressure [32]. The elastic parameters were extracted by employing the Wien2k interfaced IRelast software, which has been specifically designed for cubic structures. The cubic lattice symmetry allows the reduction of elastic constants $C_{ij}$ to only three i.e., $C_{11}$, $C_{12}$, and $C_{44}$ while other elastic parameters can be extracted from these constants. Table 2 provides an overview of the computed elastic constants along with the additional elastic parameters estimated from these constants. In cubic structures, the mechanical stability is subject to the restrictions imposed on the elastic constants which are: $C_{11} > 0$, $C_{44} > 0$, $C_{11} - C_{12} > 0$, $B > 0$, and $C_{11} + 2C_{12} > 0$ [33]. These stability criteria are met by our estimated $C_{ij}$ values, highlighting that the compounds under examination are elastically stable.

The $C_{11}$ value for NaTiF$_3$ (134.952 GPa) is quite larger than the value for NaScF$_3$ (114.195 GPa) suggesting that the NaTiF$_3$ is harder than NaScF$_3$. The appearance of microcracks is strongly associated to the elastic anisotropy of materials and has considerable implications in engineering studies. To explore the elastic anisotropy of the studied compounds, we computed the anisotropic factor (A) from the elastic constants by employing the following relationship:

$$A = \frac{2C_{44}}{[C_{11}-C_{12}]} \qquad (3)$$

For the anisotropic materials, A is lesser or greater than 1, whereas $A = 1$ designates the isotropic nature of the materials. The observed values (listed in Table 2), indicate that our compounds are highly anisotropic as reflected by their extent of variations from 1. The estimated value of $A$ is 0.308 for NaScF$_3$ and 0.038 for NaTiF$_3$, demonstrating that NaTiF$_3$ exhibits a higher degree of anisotropy. Additional elastic parameters like Bulk modulus ($B$), Poisson's ratio ($\upsilon$), Young modulus ($E$), and Shear modulus ($G$) were assessed from the observed ECs by using the following equations and the recorded values are displayed in table 2.

$$B = 1/3\,[2C_{12} + C_{11}] \tag{4}$$

$$E = \frac{9GB}{[G+3B]} \tag{5}$$

$$\upsilon = \frac{-2G+3B}{2[G+3B]} \tag{6}$$

$$G_R = \frac{5C_{44}(C_{11}-C_{12})}{3(C_{11}-C_{12})+4C_{44}} \tag{7}$$

$$G_\upsilon = \frac{1}{5}[3C_{44} - C_{12} + C_{11}] \tag{8}$$

$$G = 1/2\,[G_\upsilon + G_R] \tag{9}$$

In contrast to the bulk modulus, which reflects resistance to the fracture, the shear modulus designates resistance to plastic deformation. The bulk modulus for NaTiF$_3$ (75.695) is larger than that of NaScF$_3$ (69.599) representing that NaTiF$_3$ is more resistant to the applied external forces. NaScF$_3$ is harder and offers stronger resistance to deformation because its Shear modulus (16.910) is higher than that of NaTiF$_3$ (10.760). Young's modulus is the most reliable predictor of a material's stiffness. The computed values of Young's modulus reveal that the studied compounds are stiffer. The ductility or brittleness of a material can be judged through several factors. For example, Cauchy's pressure given by $C_{11} - C_{44}$ can be used to explore the nature of a material [34]. A positive value of Cauchy's pressure stipulates that the material is ductile while the brittleness of the material can be predicted by a negative value. The observed positive values for NaYF$_3$ (Y = Sc and Ti) illustrate that both of the studied compounds are ductile. In addition to the ductility and brittleness of a compound, the Cauchy's pressure value also uncovers the bonding nature in a material. If $C_{12} - C_{44} > 0$, the bonding in the material will be ionic, whereas negative values indicate the covalent bonding of a compound [35]. The observed positive values for NaYF$_3$ (Y = Sc and Ti) reveal that ionic bonds are present in the studied compounds.

The Pugh ratio ($B/G$) is another metric that may be used to classify whether a material is brittle or ductile. According to the Pugh criteria, a material has a brittle character if $B/G \leq 1.75$ otherwise ductile [36]. The observed $B/G$ value is 4.116 for NaScF$_3$ and 7.035 for NaTiF$_3$ demonstrating that both compounds are ductile where the ductility of NaTiF$_3$ is higher than NaScF$_3$. Another criteria, used to differentiate between the ductility and brittleness of a compound, was laid byFrantsevich et al. [37] by defining the Poissons ratio ($v$). If $v > 0.26$ the compound will be ductile; else, it will be brittle. For the studied ternary compounds, the observed Poisson's ratio are grater than the threshold value of 0.26 supporting the ductile nature which is in complete agreement with the criteria laid by Pugh as well as the Cauchy's pressure. In a nutshell, we found that the compounds under investigation are anisotropic, ductile, and reveal high toughness and fracture resistance. Based on these findings, we consider the applicability of studied compounds in various electrical technologies.

## 3.3 Electronic Properties

In this section, we exploited the TB-mBJ approximation to explore the electronic nature of NaYF$_3$ (Y = Sc and Ti) by modeling their energy band structures and the density of states (DOS). Fig. 3(a-b) and Fig. 3(c-d) illustrate the energy bands corresponding to NaScF$_3$ and NaTiF$_3$, respectively. The Fermi level (E$_F$), marked by a horizontal dashed line at 0 eV, serves as a crucial reference point. Notably, it is evident from the figures that the spin-up and spin-down channels manifest distinct band structures, thereby highlighting that these materials possess half-metallic nature. Specifically, in the spin-up configuration, both NaScF$_3$ and NaTiF$_3$ resemble metals, as the Fermi level falls within the conduction band. However, upon examination of the spin-down states of NaScF$_3$ (Figure 3(b)) and NaTiF$_3$ (Figure 3(d)), a contrasting picture emerges. From the figures, it is obvious that for spin-down configuration these compounds have wide bandgaps and depict an insulating nature. Additionally, as both the conduction and valence bands do not reach the Fermi levels, they exhibit insulating/semiconducting nature. For both compounds, in the spin-down configuration, the valence band minima align with the M symmetry point, while the conduction band maxima reside at the Γ symmetry point, thereby exhibiting an indirect (M-Γ) band nature. The observed metallic and insulating nature for the spin up and down configurations, respectively, reveals that the studied compounds have half-metallic nature.

Density of states (DOS) calculations serve as a powerful tool for elucidating the energy distribution across distinct states within a material. The metallic and non-metallic nature of a compound can be accurately grasped by calculating the density of states. To comprehend the electrical behavior of NaYF$_3$ (Y = Sc and Ti), we plotted their DOS which shows their electronic states' contributions to the conduction and valence bands (See Fig. 4). In DOS curves, the valence band exhibits negative energy values and is situated to the left of the Fermi level (denoted by a vertical dashed grey line), whereas, the region to the right signifies the conduction band featuring positive energy values. A closer examination of the DOS curves unequivocally supports the assertion of metallic properties within these materials. Fig. 4 depicts that for spin-up configurations, the major contribution to the conduction band stems from Y atoms (Sc, Ti) indicating that these elements are predominantly responsible for the observed metallic nature where a minute contribution to the metallic nature is from the F atom. Meanwhile, an analysis of the DOS curves for both compounds in the spin-down configuration reveals that the valence band is primarily influenced by the presence of Y (Sc, Ti) atoms, with a minor contribution from F atoms. This observation aligns with the ferromagnetic nature of the materials, evident from magnetic moment values of 2.00 $\mu_B$ for NaTiF$_3$ and 1.01 $\mu_B$ for NaScF$_3$.

## 3.4 Optical Properties

In this section, we explore how materials respond to incident light. The optical characteristics are mainly correlated to the dielectric function, refractive index, absorption coefficient, reflectivity, energy loss function, optical conductivity, and extinction coefficient. Herein, we report the optical characteristics of the lead-free fluoro-perovskites NaYF$_3$ (Y = Sc and Ti) for the energy values ranging from 0 to 41 eV.

3.4.1 Dielectric Function

The materials' properties are substantially impacted by their frequency-dependent dielectric function $\varepsilon(\omega)$ [38]. According to the Cohen's and Ehrenreich equation, the real and imaginary parts of the complex dielectric function represented by $\varepsilon_1(\omega)$ and $\varepsilon_2(\omega)$, respectively are related by [39,40]

$$\varepsilon(\omega) = \varepsilon_1(\omega) + i\varepsilon_2(\omega)$$

The Kramers-Kroning expressions allow us to determine these real and imaginary components of $\varepsilon(\omega)$.

$$\varepsilon_1(\omega) = 1 + \frac{2}{\pi} P \int_0^\infty \left[\frac{(\acute{\omega}^2 - \omega^2)}{\acute{\omega}\varepsilon_2(\acute{\omega})}\right]^{-1} d\acute{\omega}$$

$$\varepsilon_2(\omega) = \frac{4}{\pi\omega^2} \sum_{n\acute{n}} \frac{|P_{n\acute{n}}(k)|^2}{\nabla\omega_{n\acute{n}}(k)} ds_k$$

The real portion of the $\varepsilon(\omega)$ may be used to predict the dispersive behavior of the material while information regarding light absorption can be retrieved from the imaginary component of the complex dielectric function [41]. Additional optical characteristics, such as the absorption coefficient, refractive index, reflectivity, optical conductivity, energy loss function, and extinction coefficient can be extracted from these real and imaginary components of $\varepsilon(\omega)$.

The variation in the real part of $\varepsilon(\omega)$ measured for the energy values ranging from 0 to 41 eV is shown in Fig. 5(a). $NaScF_3$ and $NaTiF_3$ have $\varepsilon_1(\omega)$ values of 8.52 and 8.84 at zero frequency, respectively. The real part maximum approaches a value of 3.38 at 9.59 eV for $NaScF_3$ and 2.60 at 12.10 eV for $NaTiF_3$. For both compounds, the observed real part minima yield negative values, suggesting the metallic nature of the studied compounds.

In Fig. 5(b), the observed curves for the imaginary parts of $\varepsilon(\omega)$ are displayed. At zero frequency, the corresponding values of $\varepsilon_2(\omega)$ for $NaScF_3$ and $NaTiF_3$ are 11.28 and 13.19, respectively. Here the imaginary part maximum reaches a value of 3.32 at 11.52 eV for $NaScF_3$ and 2.44 at 12.31 eV for $NaTiF_3$, respectively. The imaginary portion of the complex $\varepsilon(\omega)$ signifies the material's absorption, and the studied compounds yield maximum optical absorption for the energy ranges of 9 – 14 eV and 30 – 40 eV.

3.4.2. Absorption coefficient

The material's absorption coefficient entails how these compounds might react to the incident radiations and reveal a material's capacity for absorbing incoming photons of certain energy [42]. In a nutshell, we can say that the absorption coefficient specifies how many photons a substance can absorb in a unit length. The incident photons, after getting absorbed into the materials, impart sufficient energy to the materials and thus result in the electrons transfer from the VB to CB. The following expression can be utilized to extract the absorption coefficient values from the real and imaginary parts of $\varepsilon(\omega)$.

$$I(\omega) = \sqrt{2}\omega \sqrt{[\varepsilon_1^2(\omega) + \varepsilon_2^2(\omega)]^{1/2} - \varepsilon_1(\omega)}$$

The calculated absorption coefficients of NaYF$_3$ (Y = Sc and Ti) are displayed in Fig. 5(c). It is evident from the observed spectrum that these compounds absorb energy in the ranges of 8 – 40 eV. The observed curve reveals that these materials serve as efficient light absorbers and can be exploited as potential materials for the opto-electronic devices operating in the UV region. The absorption coefficient yields maximum values of 248.21 and 287.67 for NaScF$_3$ and NaTiF$_3$, respectively, indicating that NaTiF$_3$ is a good absorber as compared to NaScF$_3$.

3.4.3. Refractive Index

The refractive index ($n(\omega)$) of a compound, which is closely associated with microscopic atomic interactions, determines how much light is refracted by that compound and plays a significant role in designing the optical components such as optical lenses, pigments, and optical thin films [43]. The refractive index can be retrieved from the dielectric function by using the following expression

$$n(\omega) = \frac{1}{\sqrt{2}} \sqrt{[\varepsilon_1^2(\omega) + \varepsilon_2^2(\omega)]^{1/2} + \varepsilon_1(\omega)}$$

The material's suitability for use in optical devices is subject to its refractive index. For example, materials with high refractive indices can be employed in photo-voltaic cells. The computed refractive indices of NaYF$_3$ (Y = Sc and Ti) are shown in Fig. 5(d). The observed spectra closely mimic the real component of complex dielectric function. The measured refractive indices of NaScF$_3$ and NaTiF$_3$ have a zero frequency limits of 3.36 and 3.52, respectively. We noticed that the maximum value of $n(\omega)_{max}$ is 1.86 at 9.62 eV for NaScF$_3$ and 1.69 at 12.12 eV for NaTiF$_3$. It is clearly visible that for both compounds the $n(\omega)_{max}$ value declines and drops to a value below 1. The observed decrease is attributed to the loss of transparency and thus the materials begin to absorb high-energy photons which are consistent with the findings shown in Fig. 5(c).

### 3.4.4. Extinction Coefficient

The extinction coefficient ($k(\omega)$) signifies the degree of light absorption by a material and can be extracted from the dielectric functions by using the following expression [44]:

$$k(\omega) = \frac{1}{\sqrt{2}}\sqrt{[\varepsilon_1^2(\omega) + \varepsilon_2^2(\omega)]^{1/2} - \varepsilon_1(\omega)}$$

The computed extinction coefficients for NaYF$_3$ (Y = Sc and Ti) are plotted in Fig. 6(a). The $k(0)$ values for NaScF$_3$ and NaTiF$_3$ are 1.67 and 1.88, respectively. The degree of absorption is minimum in the range of 1.5 – 10 eV and upsurges with further increase in energy yielding a maximum value of 1.08 at 11.58 eV and 0.82 at 12.40 eV for the NaScF$_3$ and NaTiF$_3$, respectively. For both compounds, the extinction coefficients almost remain constant for the energy ranging between 15 and 30 eV while peaking at 32.7 and 36 eV.

### 3.4.5. Optical Conductivity

The optical conductivity $\sigma(\omega)$ refers to the phenomenon in which a light of suitable frequency strikes the material's surface and as a result the electrical conduction begins. The optical conductivity can be obtained by using the following relation [44]:

$$\sigma(\omega) = \frac{2W_{cv}\hbar\omega}{\vec{E}_o^2}$$

Where, $W_{cv}$ represents the transition probability per second. The observed optical conductivity versus energy spectra for the NaYF$_3$ (Y = Sc and Ti) are presented in Fig. 6(b). The observed curves reveal that, at higher energy, NaTiF$_3$ is more optically conductive but the NaScF$_3$ exhibits more conduction at low energies. The studied compounds show optical conduction for the energy values in the range of (9 – 40) eV, where, the $\sigma_{max}(\omega)$ values of 6098 at 32.44 eV and 9078 at 35.85 eV were observed for NaScF$_3$ and NaTiF$_3$, respectively.

### 3.4.6. Energy Loss function

An electron loses energy as it travels swiftly across the material. In addition to the inner shell ionization and inter-band transition, this energy loss can also result in Plasmon and phonon excitation [45]. The static level (1.5 – 10 eV) for the compounds under investigation exhibits no energy loss, while the energy-loss peak for NaScF$_3$ starts at 12.55 eV and for NaTiF$_3$ occurs at 13.40 eV as illustrated in Fig. 6(c). The energy loss values for both materials begin to increase considerably at higher energy and yield maxima at 33.20 and 36.40 eV for the NaScF$_3$ and NaTiF$_3$, respectively.

3.4.7. Reflectivity

Reflectivity ($R(\omega)$) is one of the crucial optical features that describe how the material's surface reacts to the incident electromagnetic radiations and plays an active role in the materials characterization and its potential uses [46]. Reflectivity is one of the key ingredients to assess the material's appropriateness for shielding applications due to its anti-reflective coating. The reflectivity of a material can be extracted from its extension coefficient and refractive index using the following expression [44,47]:

$$R(\omega) = \left|\frac{n(\omega) - 1}{n(\omega) + 1}\right| = \frac{(n(\omega) - 1)^2 + k^2(\omega)}{(n(\omega) + 1)^2 + k^2(\omega)}$$

Figure 6(d) represents the computed reflectivity as a function of energy. Reflectivity at zero energy $R(0)$, for NaScF$_3$ and NaTiF$_3$, is 0.38 and 0.41, respectively. The studied compounds are greatly reflective in the energy ranges of 9 - 14 eV and 30 - 40 eV.

4. Conclusions

This study presents pioneering theoretical calculations on NaYF$_3$ (Y = Sc and Ti), using DFT, to investigate the unique composition of fluoro-perovskites in great detail. The stability of examined perovskites was validated by exploring the structural and elastic parameters, which provides assurance regarding their potential synthesis in practical applications. Both compounds exhibit ductile nature, which further highlight their less vulnerability to breaking. Furthermore, the elastic criteria unveiled that these compounds are ionic-bonded and anisotropic. The electronic properties, particularly the band profiles revealed that the investigated structures exhibit half-metallic character and thus highlighted their usefulness for applications in spin devices. For the spin-up configuration, both compounds are found to have metallic nature, whereas, for the spin-down configuration, they exhibit indirect wide band gaps in the insulating zone. The observed half-metallicity is further supported by examination of the density of states. The broad range (8 - 40 eV) absorption in the UV region reveals that these materials serve as efficient light absorbers and can be exploited as potential materials for optoelectronic devices operating in the UV region.

**Acknowledgment**

Authors wish to thanks Researchers Supporting Project Number (RSP-2023R45) at King Saud University Riyadh Saudi Arabia for financial support.

**Funding**

This research work was supported by researchers supporting Project number (RSP-2023R45) King Saud University, Riyadh, Saudi Arabia.

**Figure Captions**

**Figure 1.** The unit cell crystal structure of $NaScF_3$

**Figure 2.** The Birch-Murnaghan fitted energy versus volume for (a) $NaScF_3$ and (b) $NaTiF_3$ cubic perovskites

**Figure 3.** Computed band diagram for (a) Spin-up configuration of $NaScF_3$, (b) Spin-down configuration of $NaScF_3$, (c) Spin-up configuration of $NaTiF_3$, and (d) Spin-down configuration of $NaTiF_3$

**Figure 4.** The DOS of $NaYF_3$ (Y = Sc and Ti) computed for both spin-up and spin-down configurations

**Figure 5.** The computed (a) $\varepsilon_1(\omega)$, (b) $\varepsilon_2(\omega)$, (c) Absorption coefficient, and (d) Refractive Index for the $NaYF_3$ (Y = Sc and Ti)

**Figure 6.** The computed (a) Extinction Coefficient, (b) Optical Conductivity, (c) Energy loss function, and (d) Reflectivity for the $NaYF_3$ (Y = Sc and Ti)

**Figure. 1**

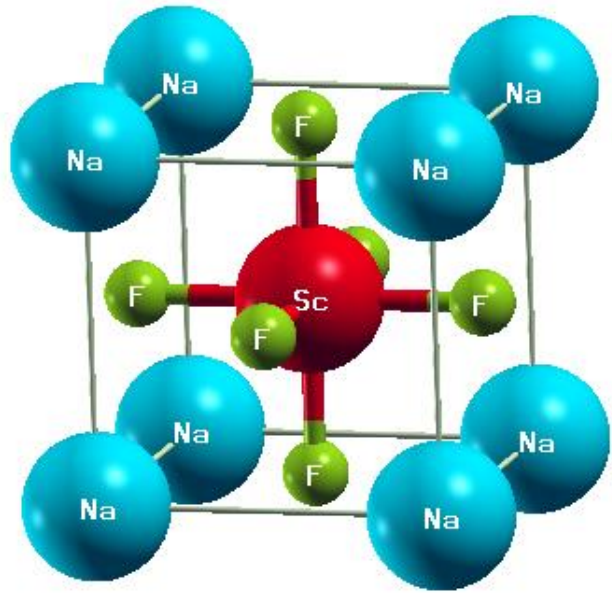

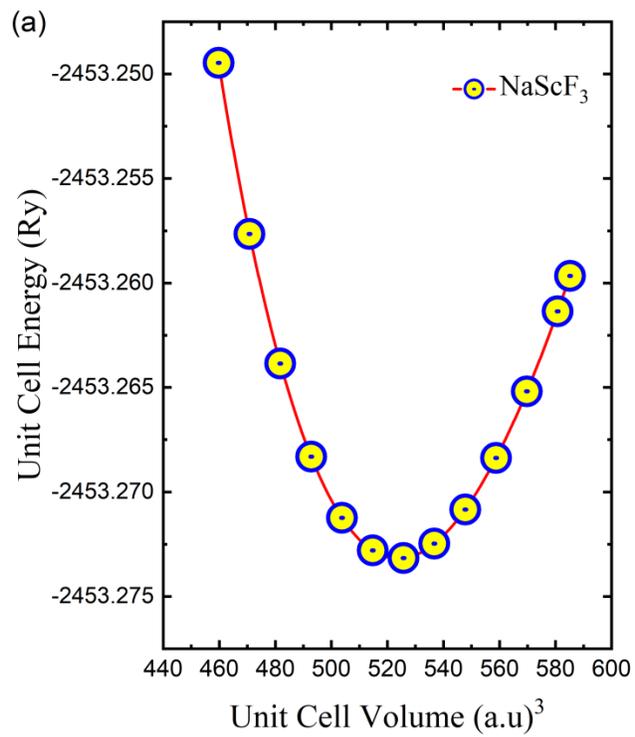

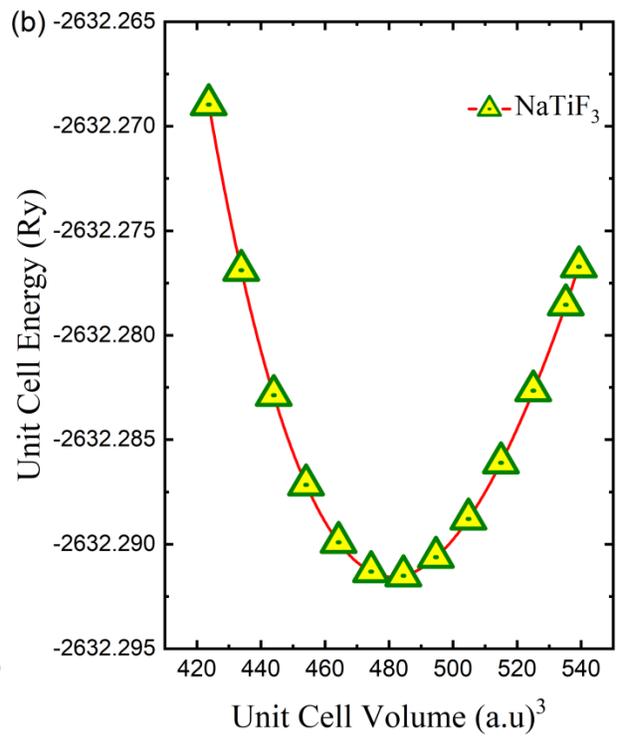

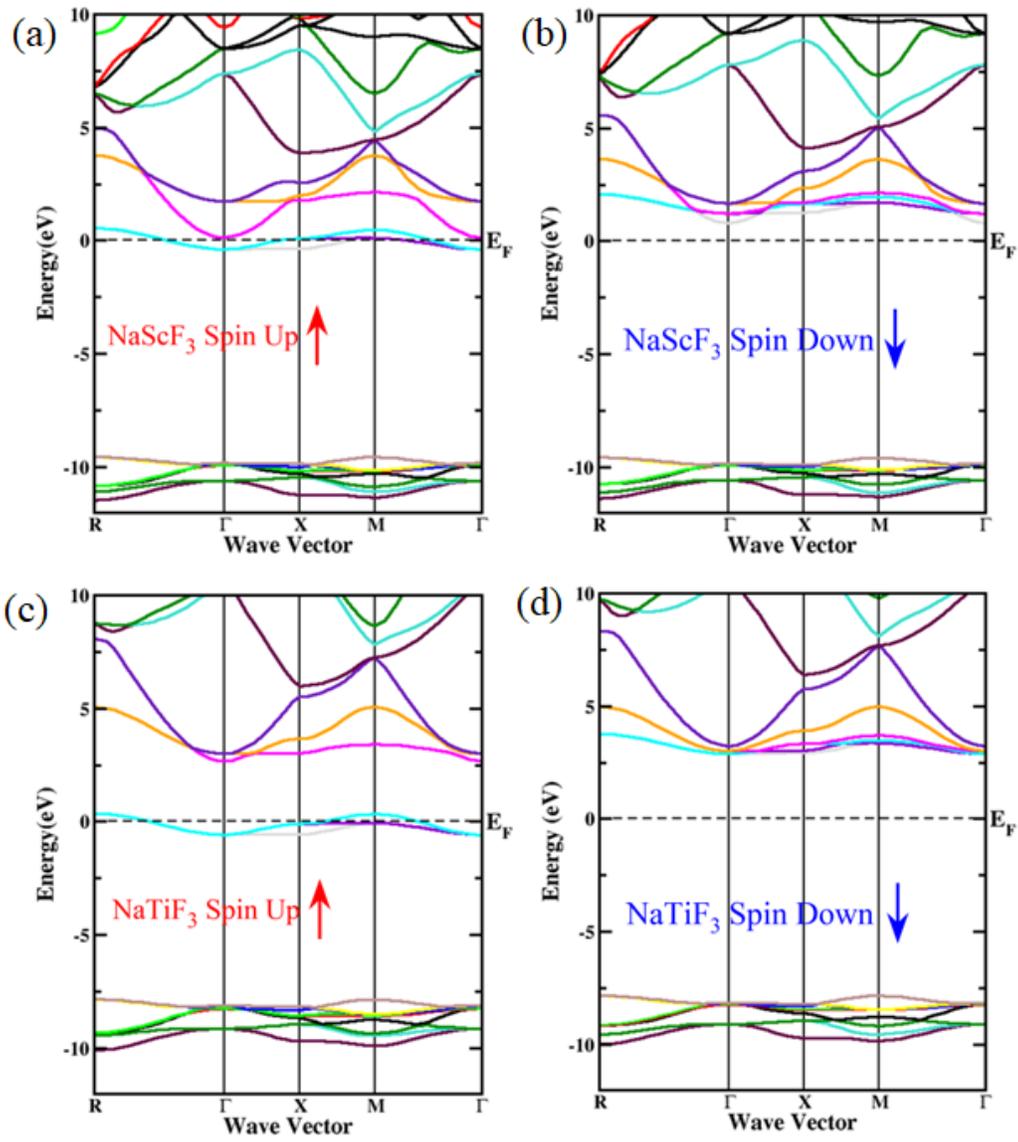

**Figure. 3**

**Figure. 4**

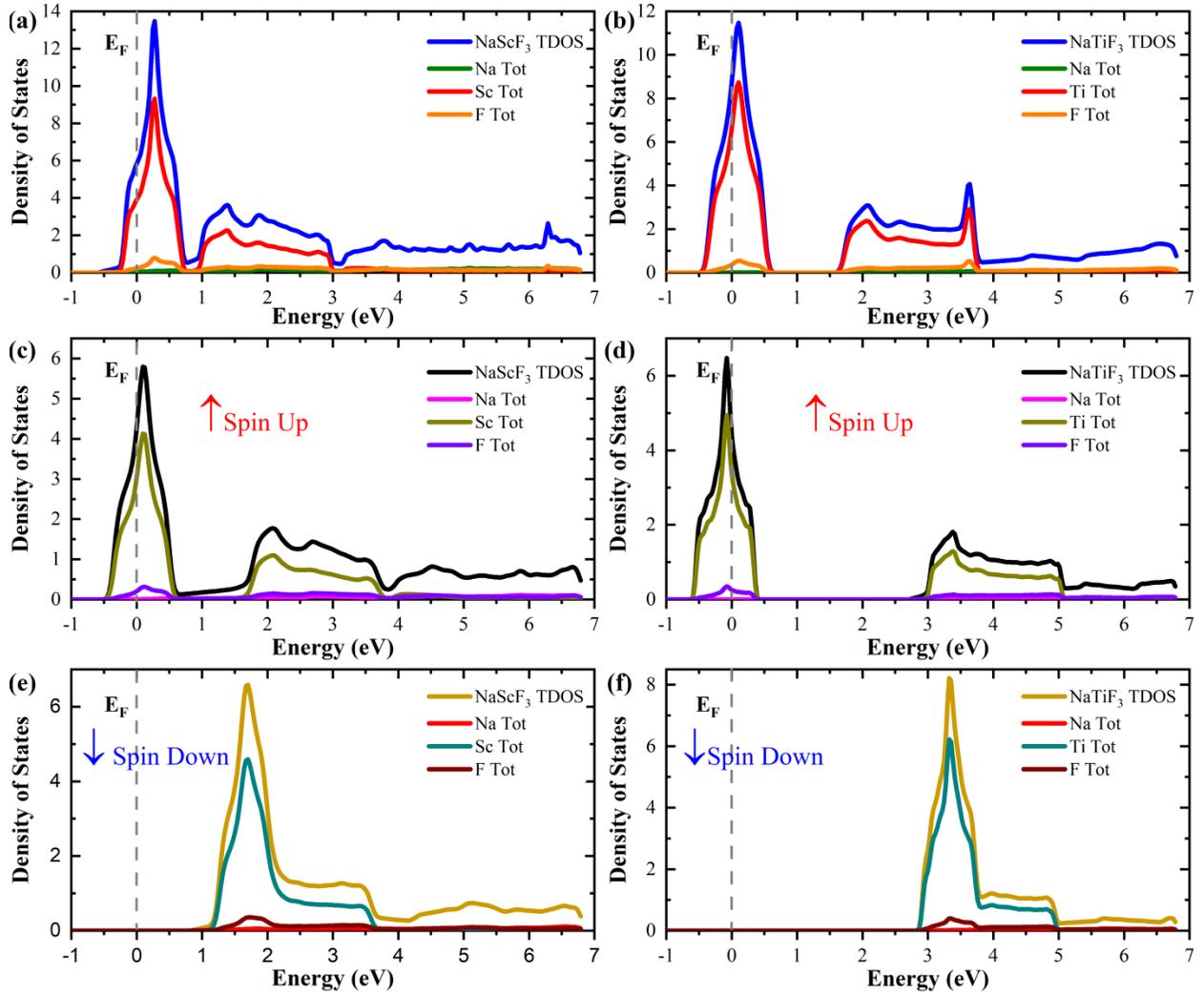

**Figure. 5**

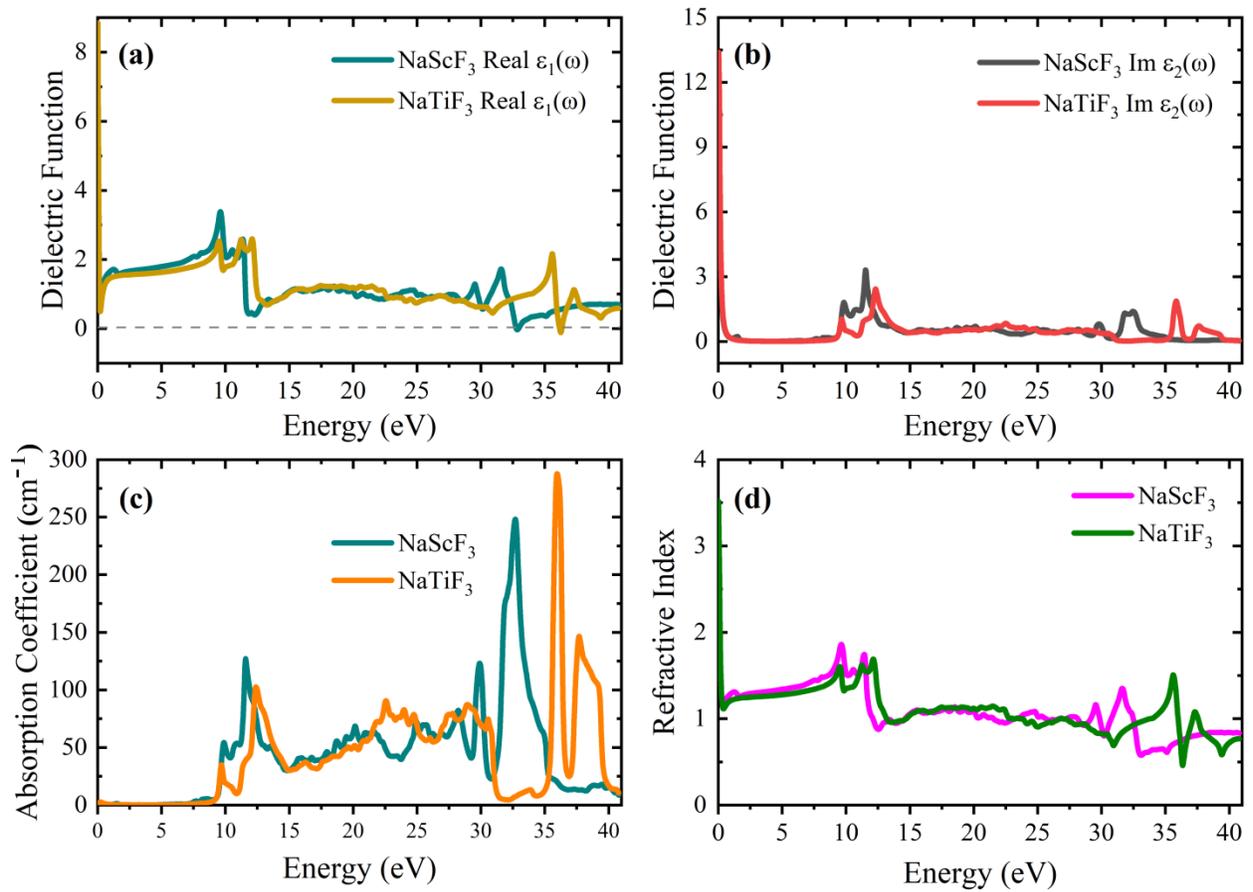

**Figure. 6**

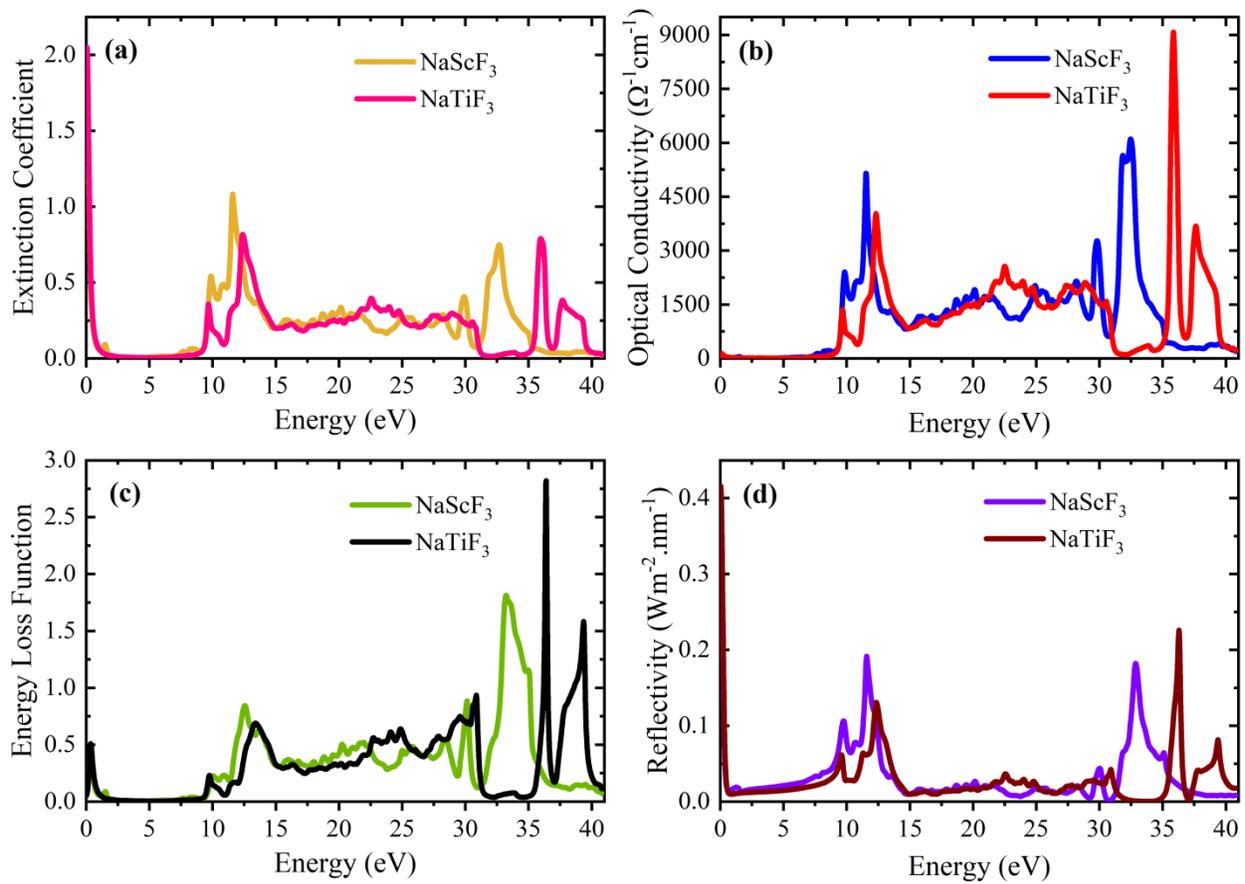

**Table Captions**

**Table 1.** Optimized lattice parameters of NaYF$_3$ (Y = Sc and Ti)

**Table 2.** The IRelast-based computed elastic parameters of NaYF$_3$ (Y = Sc and Ti)

**Table:1**

| Optimized lattice parameters | NaScF$_3$ | NaTiF$_3$ |
|---|---|---|
| $a_0(Å)$ | 4.26 | 4.15 |
| $V_0(a.u^3)$ | 523.86 | 481.16 |
| $B(GPa)$ | 68.87 | 76.72 |
| $Ḃ(GPa)$ | 4.88 | 4.52 |
| $E_0(Ry)$ | -2453.57 | -2632.30 |

**Table:2**

| Elastic Parameters | NaScF$_3$ | NaTiF$_3$ |
|---|---|---|
| $C_{11}(GPa)$ | 114.195 | 134.952 |
| $C_{12}(GPa)$ | 47.225 | 46.113 |
| $C_{44}(GPa)$ | 10.300 | 1.685 |
| $B(GPa)$ | 69.599 | 75.695 |
| $C_{11} - C_{12}(GPa)$ | 66.97 | 88.839 |
| $C_{12} - C_{44}(GPa)$ | 36.925 | 44.428 |
| $A$ | 0.308 | 0.038 |
| $E(GPa)$ | 53.689 | 52.035 |
| $v$ | 0.371 | 0.385 |
| $B/G$ | 4.116 | 7.035 |
| $G(GPa)$ | 16.910 | 10.760 |